# Efficient Computation of Metal Halide Perovskites Properties using the Extended Density Functional Tight Binding: GFN1-xTB Method

J. M. Vicent-Luna, S. Apergi, and S. Tao*

Materials Simulation and Modelling, Department of Applied Physics, Eindhoven University of Technology, 5600 MB Eindhoven, The Netherlands.
Center for Computational Energy Research, Department of Applied Physics, Eindhoven University of Technology, 5600 MB, Eindhoven, The Netherlands.

## Abstract

In recent years, metal halide perovskites (MHPs) for optoelectronic applications have attracted the attention of the scientific community due to their outstanding performance. The fundamental understanding of their physicochemical features is essential for improving their efficiency and stability. Atomistic and molecular simulations have played an essential role in the description of the optoelectronic properties and dynamical behaviour of MHPs, respectively. However, the complex interplay of the dynamical and optoelectronic properties in MHPs requires the simultaneous modelling of electrons and ions in relatively large systems, which entails a high computational cost, sometimes not affordable by the standard quantum mechanics methods, such as Density Functional Theory (DFT). Here, we explore the suitability of the recently developed Density Functional Tight Binding (DFTB) method, GFN1-xTB, for simulating MHPs with the aim of exploring an efficient alternative to DFT. The performance of GFN1-xTB for computing structural, vibrational and optoelectronic properties of several MHPs is benchmarked against experiments and DFT calculations. In general, this method produces accurate predictions for many of the properties of the studied MHPs, which are comparable to DFT and experiments. However, we also identify a few shortcomings, related to specific geometries and chemical compositions. Nevertheless, we believe that the tunability of GFN1-xTB is the key to resolving any observed issues and we propose specific targets, whose refinement will turn this method into a powerful computational tool for the study of MHPs and beyond.

## Introduction

Metal halide perovskites (MHPs) are novel semiconductors that have gained great scientific attention in the recent years due to their excellent optoelectronic properties, which make them suitable for applications such as perovskite solar cells (PSCs) and light emitting diodes.[1-5] MHPs have the chemical formula $ABX_3$, where A is a monovalent organic or inorganic cation ($Cs^+$, $CH_3NH_3^+$, and $CH(NH_2)_2^+$), B is a metal divalent cation (typically $Pb^{2+}$ or $Sn^{2+}$), and X are halide anions ($I^-$, $Br^-$, and to a lesser extent, $Cl^-$). Combining these compounds results in a semiconductor that exhibits suitable band gaps, high light absorption performance, low exciton binding energies, long carrier diffusion lengths, and high charge carrier mobility.[6-7]

In addition, MHPs exhibit a competitive fabrication cost together with a simple route to synthesize. Despite all these desirable properties, instability issues critically hamper their industrial application.[8-9]

Nowadays, many experimental and theoretical researchers are engaged in extending the understanding of the fundamental physicochemical properties of MHPs, which is crucial for increasing their stability.[10-13] Computational modeling has proven to be a valuable tool to this endeavor, since it can provide essential insights about the fundamental properties of materials that are difficult, if not impossible, to obtain experimentally. In this context, computational techniques are extremely useful in explaining the features of MHPs at the microscopic level.[12, 14-15]

There are diverse computational techniques that exhibit advantages and limitations to investigate processes at different sizes and time scales. A proper choice of the most suitable technique is often challenging because the performance of computational methods for novel or complex materials is still unknown. With this in mind, we aim to explore emerging and promising techniques that will help investigate the properties of MHPs. Density functional theory (DFT) is the golden standard of the computational methods used in materials science owing to its accuracy in predicting materials properties. In recent years, DFT calculations have been used to study many features of MHPs, such as geometrical,[16-17] optoelectronic,[16-18] and vibrational properties,[19] enthalpies of formation,[20] defect activity,[21] and ion migration,[22] among others. On the other hand, Molecular dynamics simulations (MD) based on classical force fields have proven to be useful in the study of the dynamical features of MHPs, such as ionic diffusion,[23-24] structural phase transitions,[24-25] thermal and ionic conductivities,[24] or phonon density of states.[26]

Despite the advantages of the aforementioned techniques, there are also plenty of limitations. The high computational cost of DFT calculations limits the study to small systems and short timescales. This is a considerable restriction since many of the most relevant and challenging advances in MHPs require the study of larger systems. These include alloys combining several cations, metals, and anions,[27] the effect of the concentration of vacancies and defects in the crystal,[28] the confinement of MHPs within porous materials such as silica matrices [29] or Metal-Organic Frameworks,[30] and the interface of perovskites with other materials acting as charge transport layers.[31-32] Classical simulations seem to be an "in part" solution to the above mentioned size limitations of DFT calculations, however, they suffer from other drawbacks, such as the inability to simulate electrons and chemical reactions that are essential for the description of many properties of MHPs. Besides, classical simulations need a suitable and realistic force field, which is challenging to parametrize. Therefore, an intermediate approach between DFT and classical simulations is often desired.

Semi-empirical Quantum Mechanics methods, such as density functional tight binding (DFTB) could provide an intermediate, combining the functionalities of both electronic and ionic description.[33] Traditional DFTB methods are based on simplifying the Kohn-Sham DFT total energy as a function of the electron density, using pre-computed interactions of element pairs, considerably reducing the computational cost.[34] These pair interactions as a function of the distance are tabulated and stored in the so-called Slater-Koster files. However, this parametrization lacks transferability and is limited to a number of elements, lacking parameters for the most common perovskite constituents, such as Cs, Pb, Sn and the halides.

GFN1-xTB is a new extended tight binding method, recently developed by Grimme et al., that covers all the elements of the periodic table.[35] To the best of our knowledge, GFN1-xTB is the first DFTB method that includes a proper parametrization of all the atoms existing in MHPs. This method maintains high accuracy and comprises a limited number of physically interpretable parameters that can be refined to study several key properties of given material systems. This type of DFTB method was first designed for the calculation of molecular complexes, but not for periodic systems.[35] Recently, the computation of periodic crystals via the GFN1-xTB method became possible in the Amsterdam Density Functional (ADF) suite;[36] however, its performance is still unknown. Providing its success, this extended DFTB method could play a key role in the prediction of MHPs properties, and also boost its application in materials science in general.

In this work, we investigate the effectiveness of GFN1-xTB in obtaining the main properties of MHPs. To achieve that, we analyze the energetic, structural, electronic, and vibrational properties of 18 MHPs with the formula $ABX_3$, (A = $CH_3NH_3^+$ or $MA^+$, $CH(NH_2)_2^+$ or $FA^+$, $Cs^+$; B = $Pb^{2+}$, $Sn^{2+}$; X = $I^-$, $Br^-$, $Cl^-$) in their cubic, tetragonal, and orthorhombic forms. Our results suggest that the original parametrization of GFN1-xTB describes targeted features of MHPs properly and is adequate for studying the properties of a number of MHPs. However, its performance in geometry relaxation calculations is not satisfactory, especially for the structures with lower symmetry. We find the highest limitation to be the incorrect description of the electronic properties of formamidinium cations due to the presence of complex chemical bonds, such as dynamic covalent bonds. In general, GFN1-xTB seems to be a promising method for the study of molecular and periodic systems of larger sizes, unattainable for standard DFT. With GFN1-xTB accurate results could be obtained in a fraction of the time DFT would require, however, further refinement of its parameters is required to eliminate the current limitations of this method.

## Simulation Details

The DFTB simulations presented in this work were carried out in AMS2019.3 SCM software,[36] with the implementation of the GFN1-xTB method developed by Grimme et al.[35] The GFN1-xTB Hamiltonian comprises four independent terms based on functional forms with adjustable parameters: electronic, repulsion, dispersion, and halogen-bonding terms (see refs [35, 37] for a detailed description of the method). The electronic contribution to the energy is the most relevant term of this tight binding method since it considers the electronic structure, the electrostatic, and the exchange-correlation energy. The repulsion energy is approximated by a classical expression that is independent of the electronic structure. This term is intended to correct the changes in the short-range interactions originated by the overlap of the atomic reference densities.[37] The third term, i.e., the dispersion energy, takes into account the long-range correlation effects because of the London dispersion interactions. In the GFN1-xTB method, the dispersion energy is computed by the D3 method [38] using the BJ-damping scheme.[39] Finally, the halogen-bonding term is included as a repulsive correction for the deficiencies in the description of the halogen-bonds. It is worth to mention that MHPs have a complex potential energy surface (PES) because they can be stable in different structural phases. In order to simplify the PES, we do not use the halogen bond contribution to the energy. This is justified because it is a minor correction to the energy, but it can lead to a non-continuous PES, which is not desirable for geometry optimization calculations of periodic systems.

We used the Fast Inertial Relaxation Engine [40] (FIRE) optimizer to perform all the geometry relaxations. The nuclear gradients convergence and the energy threshold for the stress tensor when optimizing lattice vectors were set to 0.001 Hartree/Å and 0.005 Hartree, repectively. Note that FIRE optimizer does not use an energy criterion convergence, but also the convergence relies on changes on forces (nuclear gradients) and stress tensor. The threshold to determine the radius of the basis functions was fixed to 0.0001, and the Coulombic interactions were computed with the Ewald summation method with a tolerance of $10^{-08}$. The grid for the K-space integration, i.e. the number of K-Points is analyzed in the first part of the Results section. The initial structures of each MHPs in their different phases were taken from the optimized structures of the previous work of Tao et al.[16]

Density Functional Theory calculations were performed using the Projector Augmented Wave (PAW) method as implemented in the Vienna Ab-Initio Simulation Package (VASP).[41-44] The electronic exchange-correlation interaction was described by the functional of Perdew, Burke, and Ernzerhof (PBE) within the generalized gradient approximation (GGA).[45] Energy and force convergence criteria of $10^{-5}$ eV and $2\times10^{-2}$ eV/Å respectively were used in all calculations, along with a kinetic energy cutoff of 500 eV and a 4×4×4 k-point grid. The D3 correction that accounts for the van der Waals interactions was employed when specified.[38] In addition, reference DFT data using PBEsol functional used for comparison were taken from a previous publication of Tao et al.[16]

## Results and discussion

### Structural properties

**K-points convergence:** The number of k-points used in quantum calculations to sample the Brillouin zone is an important parameter that influences the accuracy of the results. The use of many k-points ensures higher precision but also increases the computational cost of the simulations. A compromise between accuracy and computational cost is necessary. We first performed a set of calculations to determine how the number of k-points affects the results. By performing small deformations, i.e. isotropic expansions and compressions of the unit cell, we calculated the energy of the systems as a function of lattice parameter. The systems were confined in a fixed volume and only the ionic positions were optimized, as described in the methodology. We selected cubic $CsPbI_3$ and $MAPbI_3$ MHPs as test systems with inorganic and organic cations, respectively. We analyzed the convergence of the k-points with n = 15 k-points in each direction as reference (where the total number of k-points is $n \times n \times n$), a choice justified by our results, since the deviation of the computed energies compared with those obtained for n = 11 and 13 is almost negligible (Figure S1).

Figure 1 (a) shows the root mean squared deviation (RMSD) of the energies (Figure S1) with respect to the reference value (n = 15) as a function of the number of k-points, while Figures 1 (b) and (c) show the energies of $CsPbI_3$ and $MAPbI_3$ MHPs as a function of the lattice parameter, for a number of selected k-points. The number of k-points, as expected, affects the calculated minimum of the energy curve, but also its shape. Lower values of n lead to the prediction of smaller structures and more significant deviations on the extremes of the curves. The RMSD presented in Figure 1 (a) decreases fast up to n = 5 and plateaus for higher values. We found a good compromise between accuracy and computational cost for n = 9, and we therefore chose this number of k-points to simulate systems with lattice parameters around 6 Å, which is the standard size of the unit cell of cubic perovskites. For the larger tetragonal and orthorhombic unit cells we reduced the number of k-points accordingly.

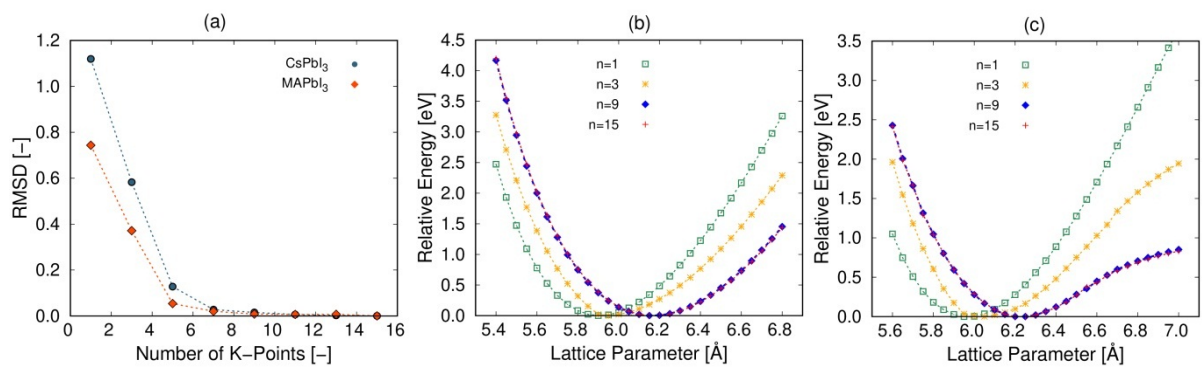

**Figure 1**. RMSD as a function of the number of k-points with n = 15 as reference (a) and k-points dependence of the energy of the GFN1-xTB optimized cubic $CsPbI_3$ (b) and $MAPbI_3$ (c) MHPs as a function of the lattice parameter. The energy of the optimal structure is set to zero. For clarity, only a set of representative k-points is presented (see Figure S1 for the complete set). *n* stands for ($n \times n \times n$) k-points in the three directions.

**Equation of states**: The ability of a computational method to describe the energy changes upon small deformation of the structures around the equilibrium is important for the description of the materials, but also for the development of potential parameters for classical simulations.[23, 25] To assess the ability of GFN1-xTB in this regard, we compared its performance with DFT in producing energy curves after isotropic distortions of the cubic MHPs. We also analyzed the effect of the dispersion energy in the simulation of MHPs crystals.

In Figure 2 the results for lead-based MHPs containing $Cs^+$ or $MA^+$ cations and $I^-$ and $Br^-$ anions are presented, with and without the dispersion energy (see the methodology section for more details) and compared to DFT data. The D3 dispersion term tends to shift the curve to lower lattice parameters, resulting in over-compression of the crystal. This correction describes the attractive part of the van der Waals interactions, which is very prominent in the molecular systems GFN1-xTB was initially developed to describe. In our case, the correction does not accurately describe the dispersion forces in the MHP crystals, therefore eliminating it from the GFN1-xTB Hamiltonian results in better agreement with the reference DFT data and experimental results. In general, the DFTB energies are in good agreement with DFT for MHPs containing inorganic $Cs^+$ cations (Figure 2 (b)). However, for $MA^+$ containing MHPs and for lattice parameters larger than the optimal, the calculated energies are slightly overestimated. This means that the GFN1-xTB total energy favors the compression of the structures around the minimum energy configuration (Figure 2 (d)).

DFT results are not unique, but instead vary depending on the selected functional and/or calculation settings. Figure S2 for instance compares the DFT lattice energies computed with the PBEsol and PBE functionals with and without the D3 dispersion term. We can see that the difference between the DFT and GFN1-xTB calculated energies is in the same range as the energy difference between two DFT functionals. We can therefore conclude that GFN1-xTB is suitable for the description of the lattice energies of these four selected MHPs. To complete the set of MHPs, Figures S3 and S4 show the relative energy data for the $CsPbCl_3$ and $MAPbCl_3$ MHPs and the same set of Sn-based MHPs. The GFN1-xTB method correctly predicts the order of the equilibrium lattice parameters following the halide order: $Cl^- < Br^- < I^-$.[16] Regarding the metal, Pb-based MHPs exhibit slightly larger unit cells than the corresponding Sn-based MHPs, as expected.[16]

Out of all the studied compositions, the largest discrepancy is found in the $FA^+$ perovskites. $MA^+$ and $FA^+$ cations are quite similar, being formed by carbon, nitrogen, and hydrogen atoms, so one would expect GFN1-xTB to perform similarly with structures containing these cations. However, as shown in Figure 3, upon compression and expansion of the unit cell the relative energies of $FAPbI_3$ and $FAPbBr_3$ exhibit an erratic behavior. Specifically, we found that small changes in the lattice parameters of the perovskite produce relatively high energy jumps, contrary to the smooth trends depicted in Figure 2. This unexpected behavior is due to the molecular structure of the $FA^+$ cations. Unlike $MA^+$, $FA^+$ cations contain a dynamic double bond between the carbon atom and one of the two attached nitrogen atoms. The presence of a double bond in a charged molecule affects the electronic configuration of the atoms, in a way that does not seem to be accounted for in the original GFN1-xTB parametrization and as a result the method fails to describe $FA^+$ containing MHPs.

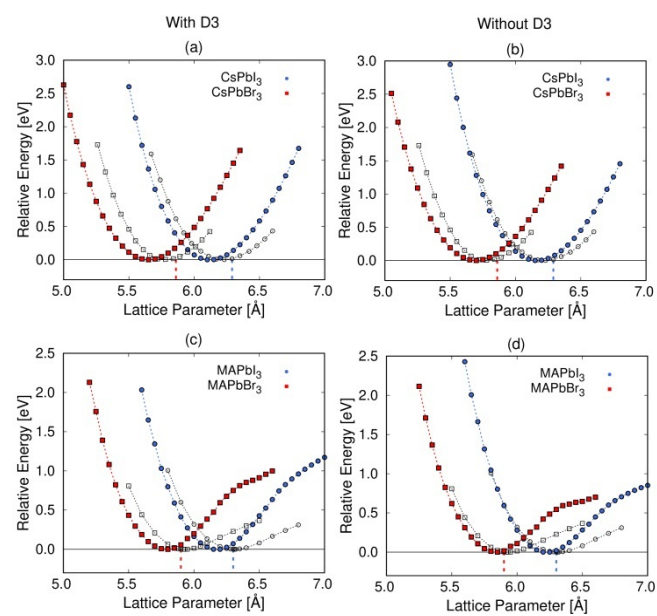

**Figure 2.** Relative GFN1-xTB energy as a function of the lattice parameter for the cubic $CsPbI_3$/$MAPbI_3$ (blue circles) and $CsPbBr_3$/$MAPbBr_3$ (red squares), with (a)/(c) and without D3 (b)/(d) dispersion corrections. DFT data (clear symbols) using PBE+D3 functional are included for comparison. The vertical dashed lines represent the experimental lattice parameters for each MHP.[46-49]

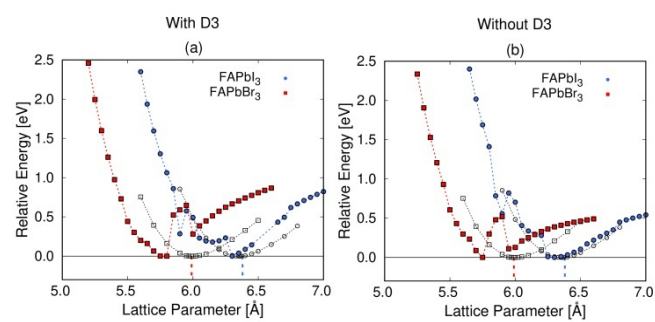

**Figure 3.** Relative energy as a function of the lattice parameter for the cubic $FAPbI_3$ (blue circles) and $FAPbBr_3$ (red squares) from GFN1-xTB with (a) and without D3 (b) dispersion corrections. DFT data (clear symbols) using PBE+D3 functional are included for comparison. The vertical dashed lines represent the experimental lattice parameters for each MHP.[16, 50]

**Organic cation rotation barrier:** Another significant property that a computational method should be able to describe accurately is the configuration of the cations within the $PbI_6$ octahedra of the MHPs. We put GFN1-xTB to the test, by calculating the energy of our systems as a function of the $MA^+$ cation rotation angle. Specifically, starting from the equilibrium configuration, we rotated the $MA^+$ cations in the unit cells of $MAPbI_3$ and $MAPbBr_3$ around the C-N axis and the energies were acquired by single point calculations (Figure 4).

We found a good agreement between GFN1-xTB and DFT, with GFN1-xTB being able to reproduce the energy barrier for the rotation of $MA^+$ cations predicted with DFT. Both methods suggest that the peak of the rotation energy barrier is around 180 degrees from the equilibrium angle. We observe slight differences in the energy values, which can be attributed to the fact that equilibrium geometries from GFN1-xTB and DFT (PBE+D3) are slightly different, and the rotation energies can only be calculated via single point calculations. It is also worth noting that for single point calculations, the D3 relative energies are the same as the ones without D3. This is because the geometry of the system does not change during the calculation, then the D3 term only contributes to the total energy with a constant value. Figure S5 shows the corresponding results for the rotation of $FA^+$ cations in $FAPbI_3$, where the inability of GNF1-xTB to properly describe $FA^+$ is manifested once more.

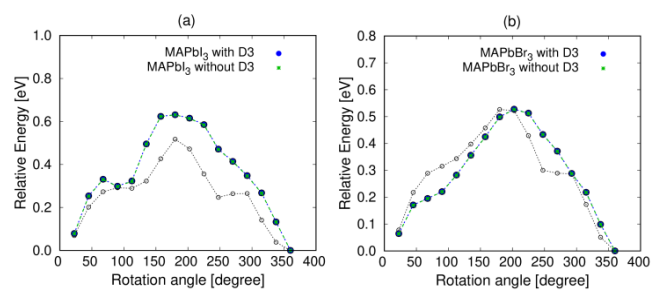

**Figure 4.** Relative GFN1-xTB energy as a function of the rotation angle of the organic $MA^+$ cations in the cubic $MAPbI_3$ (a) and $MAPbBr_3$ (b) with D3 and without D3 dispersion corrections. DFT data (open symbols) using PBE+D3 functional are included for comparison.

**Structural optimization:** All the previous results are based on calculations with the systems having a fixed volume in their cubic form. However, a successful computational method needs to be able to predict equilibrium structures through full geometry optimizations. In Figure 5 the results of the full geometry optimizations of the cubic, tetragonal, and orthorhombic phases of all MHPs studied in this work are compared with the DFT reference values. In general, GFN1-xTB tends to underestimate the lattice parameters of the MHPs, resulting in an over-compression of the material (Figure 5a). Still, most of the GFN1-xTB calculated data follow the same trend as the reference. However, a few points deviate considerably from the reference values, indicating a vast deformation of the crystal.

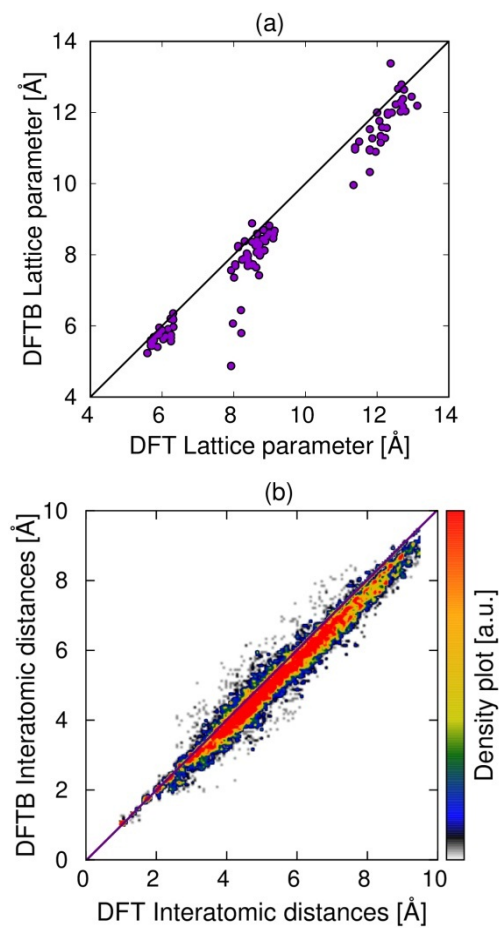

**Figure 5.** Comparison between computed DFT and GFN1-xTB lattice parameters (a) and interatomic distances (b) between all atom pairs in cubic, tetragonal, and orthorhombic MHPs. The reference DFT results were computed using the PBEsol functional without D3 dispersion corrections.[16]

To distinguish which structures and/or elements are harder for GFN1-xTB to describe, we analyzed the interatomic deviations for each system separately. Figure 6 shows the percentage of deviation of the interatomic distances by varying the A, B, and X species and the crystal shape. In general, GFN1-xTB predicts better the geometries of more symmetrical phases than the crystals with less symmetry following the order: cubic > tetragonal > orthorhombic. The higher distortion is observed for orthorhombic phases reaching values of deviation up to 20-30% from the reference data, while cubic structures are predicted with a maximum error lower than 10-15%.

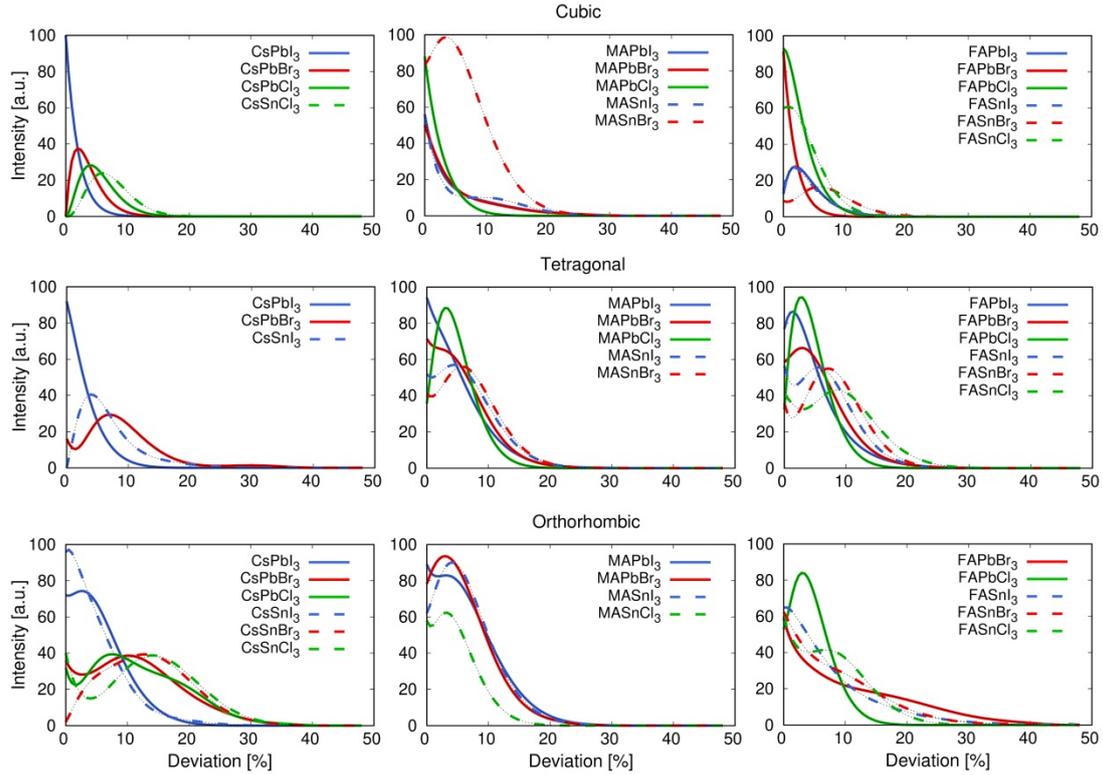

**Figure 6**. Histogram of the percentage of deviation of interatomic distances between all atom pairs calculated with GFN1-xTB with respect to the reference DFT values in cubic, tetragonal, and orthorhombic MHPs. The reference DFT results were computed using the PBEsol functional without D3 dispersion corrections.[16]

MHPs containing $Cs^+$ and $I^-$ seem to be better predicted by GFN1-xTB, while the method also performs slightly better for Pb-based MHPs than their respective Sn-based MHPs. We can conclude that the most influential factor for the performance of GFN1-xTB is the geometry of the studied system. The method describes the simplest and high-symmetry cubic geometries better than the more complex and distorted orthorhombic phases. This can be related to the fact that the GFN1-xTB Hamiltonian is based on the interatomic distances between pairs of atoms, which are more uniformly distributed in more symmetrical systems.

**Electronic properties**

**Band structure:** Another advantage of the DFTB methods over classical simulations is the ability to describe electronic properties. To benchmark the performance of GFN1-xTB in the prediction of the electronic properties of MHPs, we compared the GFN1-xTB calculated electronic band structure of the most typical $CsPbI_3$ and $MAPbI_3$ MHPs with the respective DFT band structures. As can be seen in Figure 7, there is excellent agreement between GFN1-xTB and DFT, with the same observed trends for the more important bands, i.e., those closer to the conduction and the valence bands. Both MHPs exhibit a direct band gap at the Γ point with values of 1.91 and 1.93 eV for $CsPbI_3$ and 1.61 and 1.66 eV $MAPbI_3$ obtained with DFT and GFN1-xTB calculations, respectively.

We also confirmed that these predictions are in line with the experimental observations, by calculating the band gaps of the $CsBX_3$ and $MABX_3$ MHPs. From the geometrical analysis, we know that some of the GFN1-xTB optimized systems can suffer a considerable structural distortion. To account for these deformations and their effect on the calculated band gaps we compared the band structures of the systems previously optimized with DFT (DFTB-DFT-opt) to those optimized with GFN1-xTB (DFTB-DFTB-opt). The results are presented in Figure 8, together with the experimental data reported by Tao et al.[16] Most of the calculated band gaps are very close to the experimental and only a few deviate, with the largest differences observed for the $CsPbCl_3$ and $CsSnCl_3$ MHPs. GFN1-xTB also predicts the correct behavior of the band gap evolution when changing the halide anion, i.e., the band gap increases as the size of the anion decreases. We can also observe that the GFN1-xTB optimization worsens the agreement with experiments but still predicts the correct tendency. It is worth mentioning that GFN1-xTB can predict the band gaps of the perovskites similarly to more expensive DFT calculations reported in the literature.[18] Simple DFT generally overestimates band gaps, however, not taking into account relativistic effects, also leads to band gap predictions comparable to the experiments, due to error cancelation. Nevertheless, our results indicate that GFN1-xTB is also suitable for the accurate prediction of the electronic properties of MHPs.

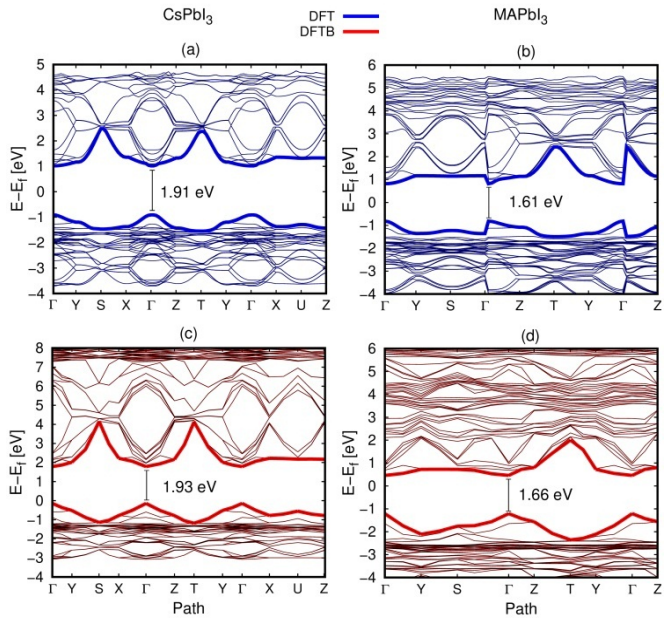

**Figure 7.** Calculated (PBE+D3) DFT (blue) and GFN1-xTB (red) band structure for the orthorhombic CsPbI$_3$ (a), (c) and tetragonal MAPbI$_3$ (b), (d).

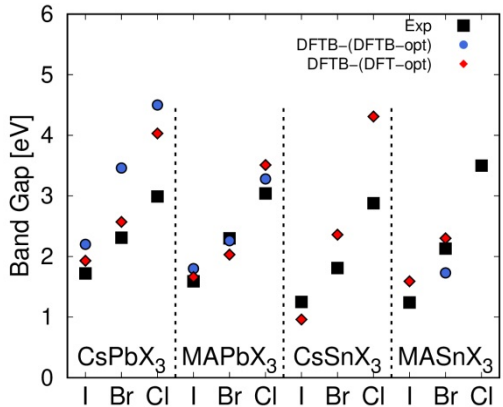

**Figure 8.** Comparison between experimental (black squares) and computed GFN1-xTB band gaps for the orthorhombic CsBX$_3$ and tetragonal MABX$_3$ (B = Pb, Sn and X = I, Br, Cl). The structures were previously optimized with GFN1-xTB (blue circles) and DFT (red diamonds) using PBEsol functional. Experimental values are taken from Tao et al.[16]

**Density of states:** To analyze the electronic properties of the MHPs in more detail, we computed the contribution of each species to the electronic density of states. In Figure 9 the GFN1-xTB calculated partial density of states (DOS) for the DFT optimized CsPbI$_3$, MAPbI$_3$, and FAPbI$_3$ is compared to its DFT counterpart. In general, GFN1-xTB gives similar results to DFT, at least around the band gap, there are however some discrepancies, with some peaks deviating up to 2 eV. The performance of GFN1-xTB is generally acceptable, but with two remarkable exceptions being the absence of peaks for Cs$^+$ around -9 and -14 eV in CsPbI$_3$, and a systematic shit to higher energies of the FA$^+$ partial DOS. The latter deviation produces a peak within the band gap of FAPbI$_3$, close to the valence band maximum that hinders the estimation of a reliable band gap value. This contrasts with the MA$^+$ cation, for which the GFN1-xTB calculated PDOS aligns well with the reference. The problem with the description of the electronic behavior of FA$^+$ seems to be in line with the erroneous description of the relative energies depicted in Figure 3.

Figure S6 shows the DOS before and after the structures have been optimized with the GFN1-xTB method. We can see that the structural changes caused by the full geometry optimization (Figures 5 and 6) do not significantly influence the general behavior of the electronic DOS, however, small energy displacements of the DOS peaks can be observed. These shifts are responsible for the differences in the computed band gaps depicted in Figure 8. It is worth mentioning that a proper optimization with GFN1-xTB before computing the electronic properties does not solve the incorrect description of the DOS of FAPbI$_3$.

**Vibrational properties**

The last examination conducted in this work involves the calculation of the vibrational properties of MHPs, which, despite not being essential to the performance of PSCs, are a fundamental piece for the study of the MHP stability and the differences between the various structural phases.[19, 46]

In Figure 10 the GFN1-xTB calculated phonon dispersion for the cubic CsPbI$_3$ and CsPbBr$_3$ is presented and compared to its DFT counterpart reported in the literature.[51-52] Certain agreement is observed between the two methods, with most of the identified vibrational modes in both MHPs fluctuating around the same energies (or frequencies). Remarkably, GFN1-xTB accurately predicts the existence of imaginary acoustic modes (negative), also known as soft modes, at the M and R points, which is a common aspect of the cubic MHPs, indicative of the dynamical instabilities of the structures. Similarly, GFN1-xTB predicts the disappearance of the imaginary modes, when replacing the cubic with the more stable orthorhombic phases (Figure S7), in concordance with the data reported in previous works.[46]

In addition to the Cs-based perovskites, we also found a similarly satisfactory performance for MA-based MHPs. Figure S8 shows the phonon dispersion for the cubic and orthorhombic MAPbI$_3$, where once more the expected suppression of the imaginary modes in the most stable phase is observed. Specifically, the phonon dispersion of the cubic MAPbI$_3$ exhibits negative modes at M and R, while for the orthorhombic phase, the lowest phonon modes have zero frequency at the gamma point, in good agreement with the data reported by Walsh et al.[19] These results suggest that GFN1-xTB is a valid tool for the efficient and accurate description of the MHPs vibrational properties.

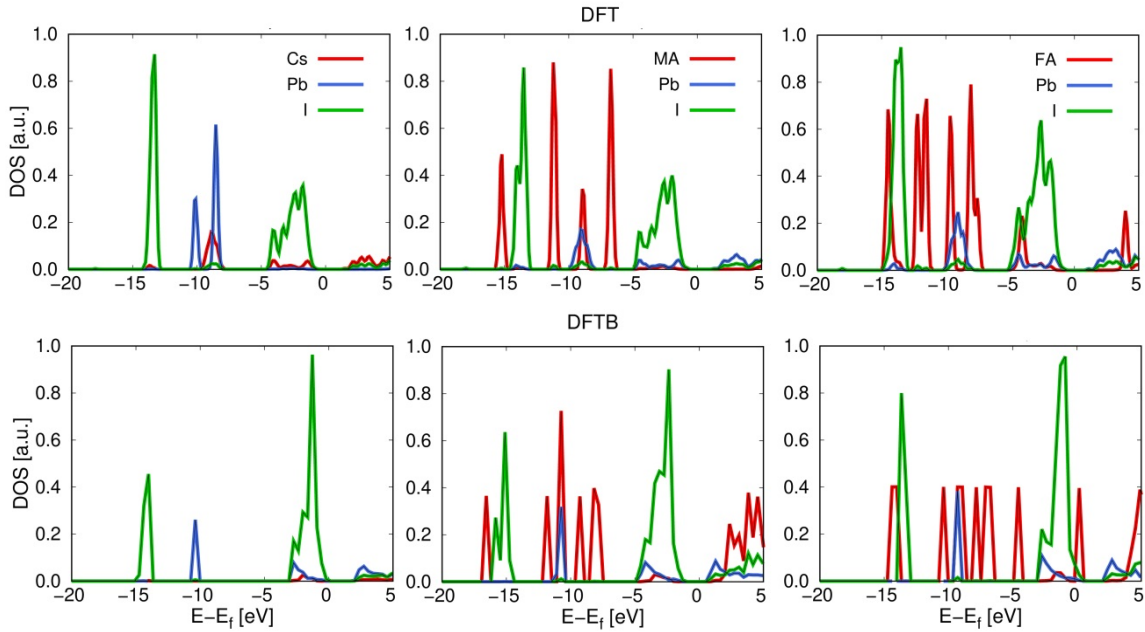

**Figure 9.** Partial density of states for the orthorhombic CsPbI$_3$, and tetragonal MAPbI$_3$ and FAPbI$_3$ computed with DFT (top) and GFN1-xTB (bottom).

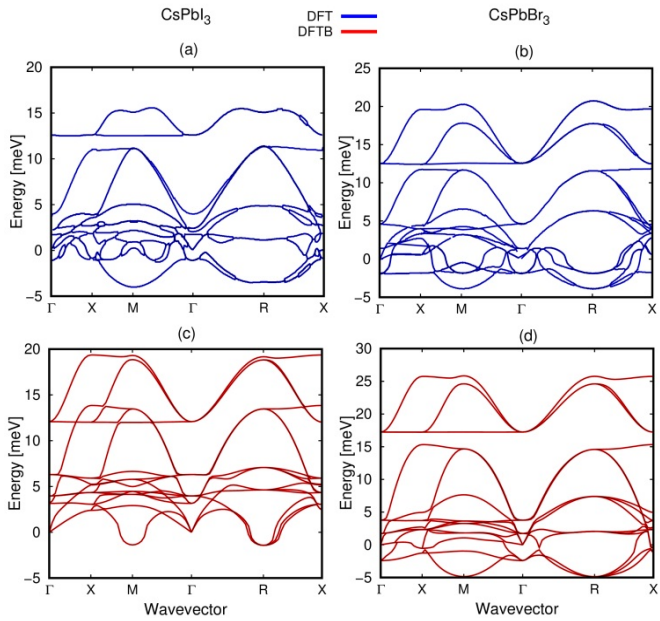

**Figure 10.** Phonon dispersion of the cubic CsPbI$_3$ (a), (c) and CsPbBr$_3$ (b), (d) computed with GFN1-xTB (red lines). Reference DFT data (blue lines) were taken from the literature.[51-52]

## Conclusions

This work provides a comprehensive overview of the performance of the semi-empirical GFN1-xTB tight binding method for the study of MHPs. Our analysis suggests that this method is suitable for the description of a variety of properties of the most common MHPs with reasonable accuracy. Such properties are 1) energetic and geometrical properties such as equations of state, rotation energy barriers of organic cations, and geometrical relaxation; 2) electronic properties such as band structures, band gaps, and partial density of states; and 3) vibrational properties such as phonon dispersions on MHPs.

Despite its general effectiveness, GFN1-xTB has some shortcomings that do not yet allow for large scale calculations of specific material properties or particular chemical compositions. Two are the main limitations we identified. The first one is the undesirable structural distortion of certain structures after a full geometry optimization. In this regard, orthorhombic phases can deviate up to 20-30% from the reference data, in contrast to the cubic crystals that show a maximum deviation lower than 10-15%. The second limitation is the inaccurate description of charged molecules with double or triple bonds between their atoms, such as FA$^+$ cations, which extends to the description of their electronic behavior.

The tunability of GFN1-xTB provides us with the ability to overcome the mentioned deficiencies. The GFN1-xTB

Hamiltonian contains various independent terms (electronic, repulsive, dispersion, and halogen-bonding terms) based on adjustable parameters that can be fitted to improve the quality of the results. Future work should focus on refining the repulsive potential parameters to avoid the observed reduction of interatomic distances and achieve the prediction of more accurate geometries. Modifying the electronic term parameters so that double and triple bonds in charged systems are properly accounted for is also necessary. DFT derived data can serve as training sets to improve the accuracy of the GFN1-xTB predictions. With further work on this line, we believe that the GFN1-xTB method can become a powerful tool to simulate not only MHPs but also other systems in the area of materials science and beyond.

## Conflicts of interest

There are no conflicts to declare

## Author Information

Corresponding Author

*E-mail: s.x.tao@tue.nl

## Acknowledgements

J.M.V.L. and S.A. acknowledge funding support from NWO (Netherlands Organization for Scientific Research) START-UP from the Netherlands. S.T. acknowledges funding by the Computational Sciences for Energy Research (CSER) tenure track program of Shell and NWO (Project No. 15CST04-2) as well as NWO START-UP from the Netherlands.

# SUPPORTING INFORMATION

## Efficient Computation of Metal Halide Perovskites Properties using the Extended Density Functional Tight Binding: GFN1-xTB Method

J. M. Vicent-Luna, S. Apergi, and S. Tao[*]

Materials Simulation and Modelling, Department of Applied Physics, Eindhoven University of Technology, 5600 MB Eindhoven, The Netherlands.
Center for Computational Energy Research, Department of Applied Physics, Eindhoven University of Technology, 5600 MB, Eindhoven, The Netherlands.

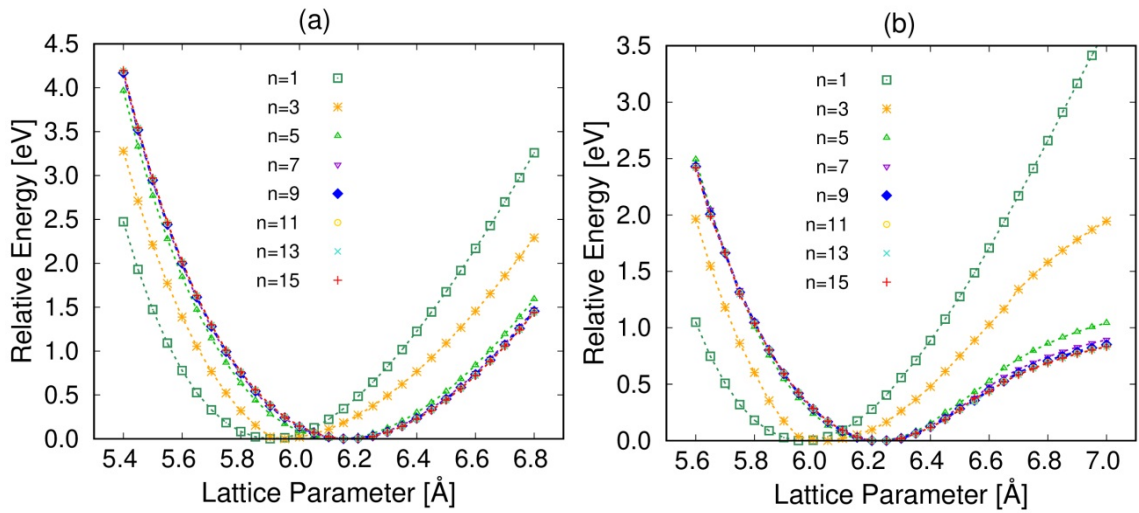

**Figure S1**. K-Points dependence on the relative energy of the GFN1-xTB optimized cubic CsPbI$_3$ (a) and MAPbI$_3$ (b) MHPs as a function of the lattice parameter. The number of K-Points (*n*) stand for (*n x n x n*) K-Points in the three directions.

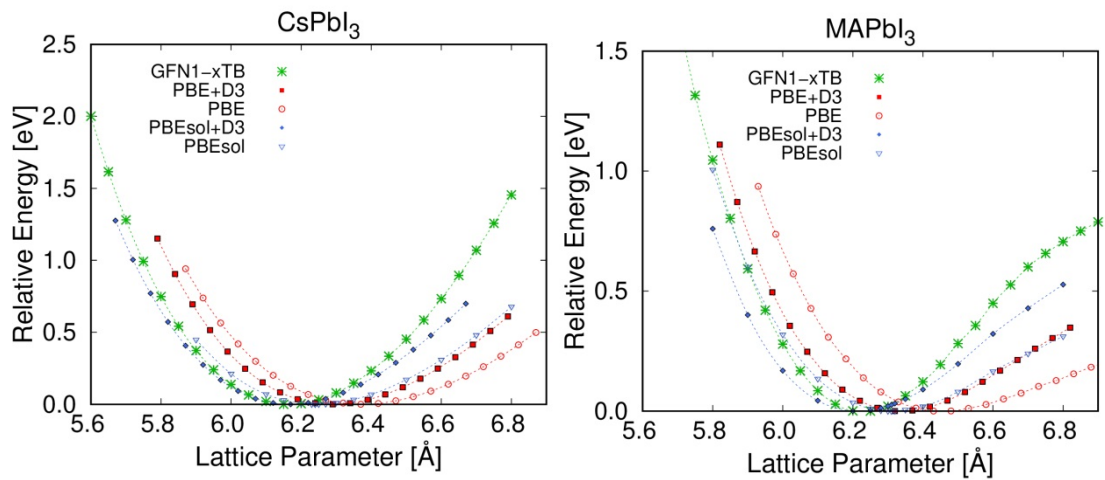

**Figure S2**. Relative energy as a function of the lattice parameter of cubic CsPbI$_3$/MAPbI$_3$ from DFT calculations with PBE and PBEsol functionals and including D3 and without including D3 dispersion corrections.

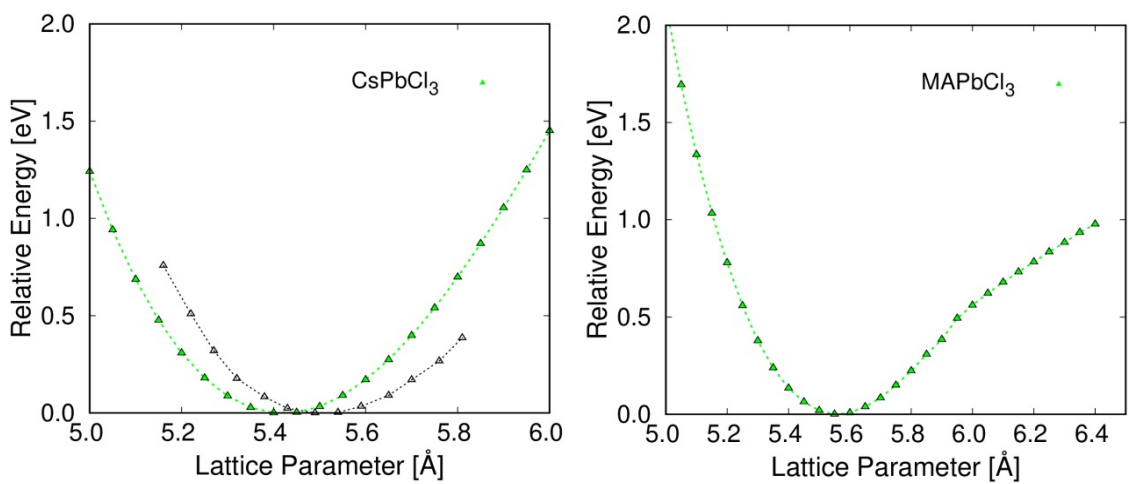

**Figure S3**. Relative energy as a function of the lattice parameter of cubic CsPbCl$_3$/MAPbCl$_3$ from GFN1-xTB without including D3 dispersion corrections.

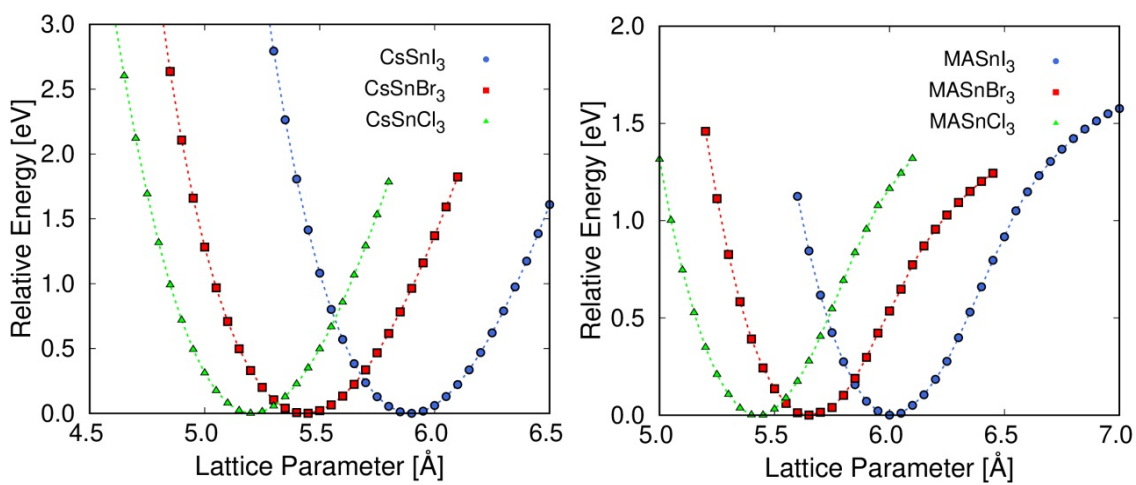

**Figure S4**. Relative energy as a function of the lattice parameter of cubic CsSnX$_3$/MASnCl$_3$ from GFN1-xTB without including D3 dispersion corrections.

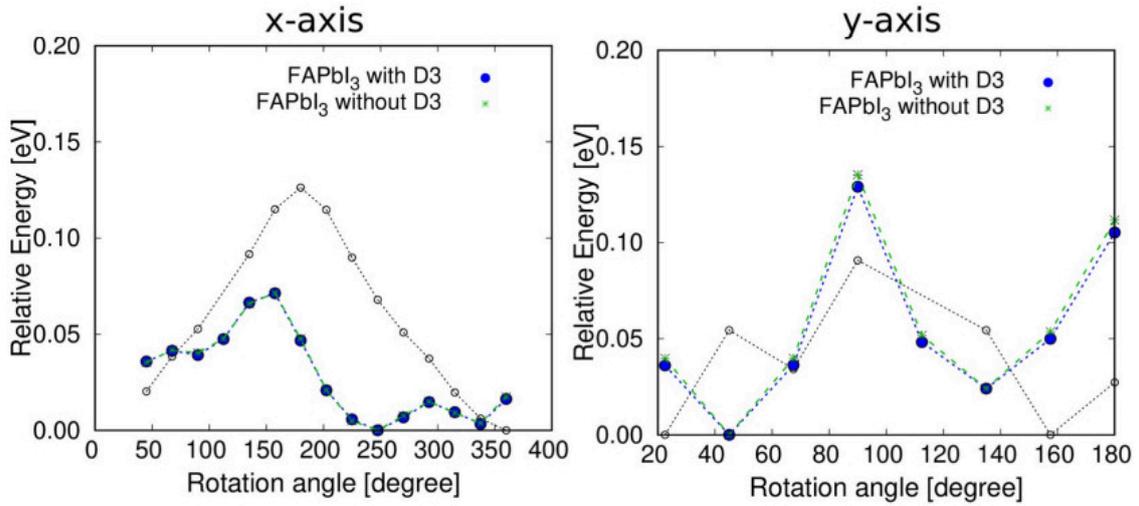

**Figure S5**. Relative energy as a function of the rotation angle of organic FA cations in cubic FAPbI$_3$ from GFN1-xTB including D3 and without including D3 dispersion corrections. DFT data (open symbols) using PBE+D3 functional is included for comparison.

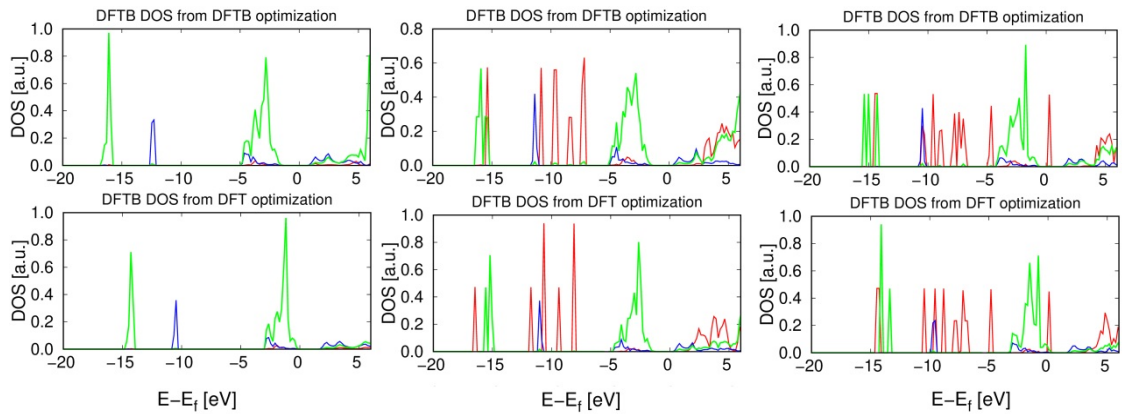

**Figure S6**. Partial density of states for orthorhombic CsPbI$_3$ and tetragonal MAPbI$_3$ and FAPbI$_3$ computed with GFN1-xTB after full geometry optimization with GFN1-xTB (top) and DFT PBEsol (bottom).

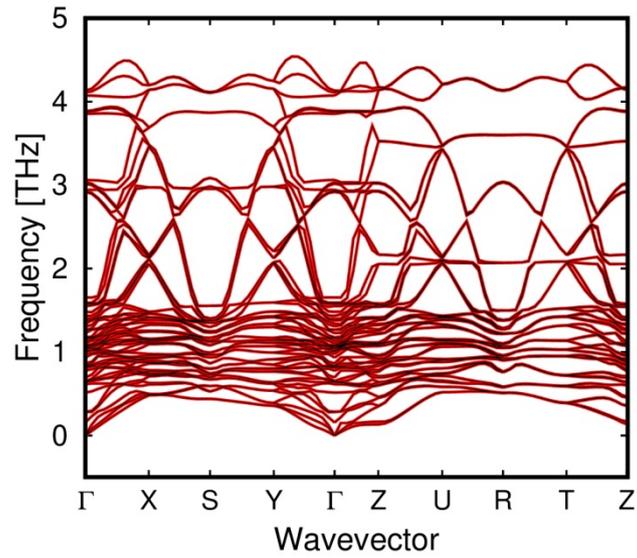

**Figure S7**. Phonon dispersion of the orthorhombic phase of CsPbI$_3$ computed with GFN1-xTB.

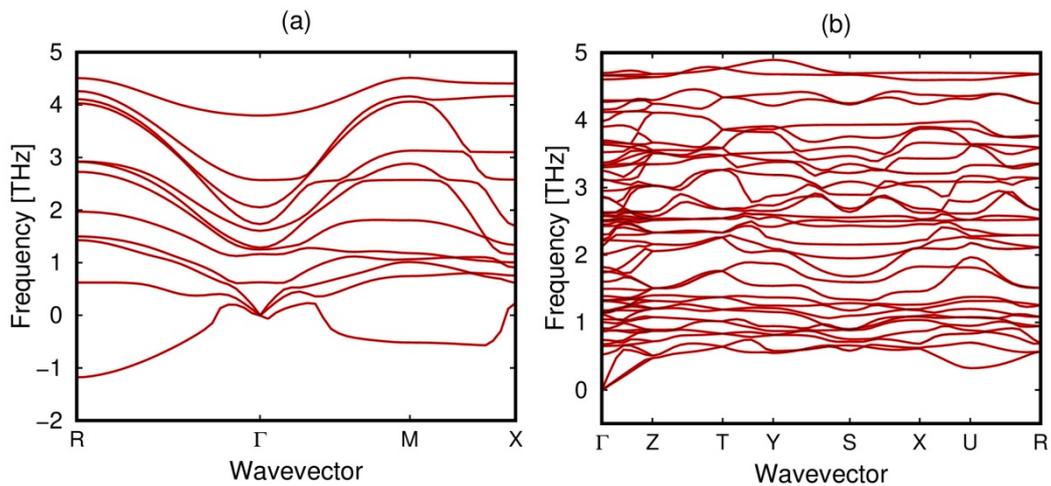

**Figure S8**. Phonon dispersion of the cubic (a) and orthorhombic (b) phases of MAPbI$_3$ computed with GFN1-xTB. Note that the results are presented in frequencies for a better comparison with the DFT values reported by Walsh et al.[1]

[1] Whalley, L. D.; Skelton, J. M.; Frost, J. M.; Walsh, A. Phonon anharmonicity, lifetimes, and thermal transport in CH$_3$NH$_3$PbI$_3$ from many-body perturbation theory. *Physical Review B* **2016**, *94* (22), 220301, DOI: 10.1103/PhysRevB.94.220301.